\documentclass[published]{JHEP3} % 10pt is ignored!
\JHEP{00(2010)000}
\JHEPspecialurl{http://jhep.sissa.it/JOURNAL/JHEP3.tar.gz}
\usepackage{epsfig,multicol,bbm}

\newcommand\fverb{\setbox\fverbbox=\hbox\bgroup\verb}
\newcommand\fverbdo{\egroup\medskip\noindent%
                        \fbox{\unhbox\fverbbox}\ }
\newcommand\fverbit{\egroup\item[\fbox{\unhbox\fverbbox}]}
\newbox\fverbbox

\newcommand{\WMAP}{{\it{WMAP }}}
\newcommand{\PLANCK}{{\it{PLANCK }}}

\newcommand{\muK}{{\mu\mathrm{K}}}
\newcommand{\fpr}[1]{\left( #1\right)}

\newcommand{\fsq}[1]{\left[ #1\right]}

\newcommand{\id}{{\mathrm{d}}}

\newcommand{\bfr}{{\bf{r}}}
\newcommand{\fabs}[1]{\left| #1\right|}

\title{Probing Cosmic Strings with Satellite CMB measurements}

\author{E. Jeong\\
        SISSA, Via Bonomea, 265, 34136, Trieste, Italy\\
        E-mail: \email{ehjeong@sissa.it}}
\author{Carlo Baccigalupi\\
        SISSA, Via Bonomea, 265, 34136, Trieste, Italy\\
        INAF-Osservatorio Astronomico di Trieste, Via G.B Tiepolo 11,I-34131 Trieste, Italy\\
        INFN/National Institute for Nuclear Physics, Via Valerio 2, I-34127 Trieste, Italy\\
        E-mail: \email{bacci@sissa.it}}
\author{G. F. Smoot\\
      Department of Physics, University of California, Berkeley, CA 94720, USA\\
      Lawrence Berkeley National Laboratory, Berkeley, CA 94720, USA\\
      Institute for the Early Universe and Department of Physics, Ewha Womans University, 120-750, Seoul, South Korea\\
      Chaire Blaise Pascal, Universite Paris, Denis Diderot\\
      E-mail: \email{gfsmoot@lbl.gov}}
\received{\today}               %%
\accepted{\today}               %% These are for published papers.

\abstract
{We study the problem of searching for cosmic string signal patterns in the
present high resolution and high sensitivity observations of the
Cosmic Microwave Background (CMB). This article discusses a technique capable
of recognizing Kaiser-Stebbins effect signatures in total intensity anisotropy
maps from isolated strings. We derive the statistical distributions of 
null detections from purely Gaussian fluctuations and instrumental performances 
of the operating satellites, and show that the biggest factor that produces 
confusion is represented by the acoustic oscillation features of the scale 
comparable to the size of horizon at recombination. Simulations show that the 
distribution of null detections converges to a $\chi^2$ distribution, 
with detectability threshold at $99\%$ confidence level corresponding to 
a string induced step signal with an amplitude of about 100 $\muK$ which 
corresponds to a limit of roughly $G\mu \sim 1.5\times 10^{-6}$. 
We implement simulations for deriving the statistics of spurious 
detections caused by extra-Galactic and Galactic foregrounds. For diffuse 
Galactic foregrounds, which represents the dominant source of contamination, 
we construct sky masks outlining the available region of the sky where 
the Galactic confusion is sub-dominant, specializing our analysis to the case 
represented by the frequency coverage and nominal sensitivity and resolution of 
the Planck experiment. As for other CMB measurements, the maximum 
available area, corresponding to 7\%, is reached where the foreground emission 
is expected to be minimum, in the 70-100 GHz interval.}


\begin{document}

\section{Introduction}
Current theories of particle physics predict spontaneous symmetry breaking
in the early Universe, with the consequent creation of coherent structures
in the spatial distribution of the fields involved in the process; these
structures, known as topological defects, represent genuine tracers and unique
remnants of those processes, occurring at energies which are not achievable
with the current terrestrial particle colliders, see \cite{copelandkibble}
and references therein. Among the most known and studied relics of this kind,
cosmic strings correspond to an uni-dimensional region in which a (scalar)
field is trapped by true minima of its potential, storing energy in that
region. The energy density is parametrized by the dimensionless quantity
$G\mu$ where $G$ is the Newton's constant, and $\mu$ the string tension, see
\cite{copelandkibble} for review.\\
Years ago, while it was still unclear if strings could dominate the
structure formation process, their signal was searched in the main statistical
indicators of structure formation itself, like the power spectra of
anisotropies in the Cosmic Microwave Background (CMB) or Large Scale Structure
(LSS). The conclusion of these studies was that strings should contribute to
cosmic structure formation by no more than a few percent
\cite{pogosian.et.al,wu,fraisse2,daviskibble,wyman.et.al,seljakslosar}.
Indeed, the evidence for coherent acoustic oscillations in the CMB and LSS
power spectra, dominated by density fluctuations, favored the scenario of
isentropic or adiabatic Gaussian fluctuations like predicted in the simplest
inflationary scenarios, and indicated that cosmic strings, if any, played a
minor role in providing the initial conditions, in a statistical sense, to the
Universe we see today. Early in their study, it was evident that strings 
produced perturbations that were not coherent as the observed acoustic 
oscillations and that they tend to produce equal power in scalar (density), 
vector (vorticity) and tensor (gravitational waves) perturbation modes. 
Recently, \cite{seljakslosar} proposed to look at sub-dominant, but most 
interesting components of the power spectrum of CMB anisotropies, namely the B 
modes activated by cosmological gravitational waves, to constrain cosmic 
strings.\\
The hypothesis is then that these objects are washed away by inflation itself
motivating the direct search of the signal from isolated and rare cosmic
strings. Among these studies, those on CMB are based on the well known 
Kaiser-Stebbins effect \cite{KaiserStebbins} resulting in a step like feature 
in the CMB temperature caused by a string which is orthogonal and moving 
relativistically with respect to the line of sight. These studies produce 
upper limits on the abundance of string in our own observable Universe of 
course, but more quantitatively on the string density parameter $G\mu$
\cite{hindmarsh1,perivolaropoulossimatos,jeongsmoot1,lowright,pogosian.et.al2,kuijken.et.al,gasparini.et.al,christiansen.et.al,morganson.et.al}.
On the other hand, studies in the optical band look for strong lensing events,
in which the strings bend the light from a distant Galaxy, causing a mirror
image of that\cite{vilenkinshellard}.\\
In this paper, we specifically search for signals imprinted on CMB 
temperature observations by long cosmic strings of cosmological scale, which 
was predicted by theory and also supported by simulations\footnote{http://www.damtp.cam.ac.uk/research/gr/public/cs\_evol.html}. 
The long strings between the last scattering surface and us would be 
back-lighted by CMB photons and leave signatures in CMB temperature map via the 
Kaiser-Stebbins effect. 
While the works mentioned above are dealing with the effects of cosmic 
string network on statistical observables, this work aims at the 
direct search of signals produced by isolated cosmic strings left on CMB 
temperature anisotropy maps, having a coherence length corresponding to 
at least the angle subtended by an Hubble horizon at decoupling.
The Planck survey\footnote{www.rssd.esa.int/planck} of CMB temperature and 
polarization anisotropies is ongoing \cite{mandolesi.et.al}, promising a leap 
forward in many research fields, including constraints on cosmic strings
abundance and tensions. The unprecedented frequency coverage, sensitivity
and angular resolution requires a dedicated discussion targeted to
the applicability and expectations from Planck direct searches of cosmic
strings. In this paper we work out two missing issues concerning the 
direct search of KS effects of cosmic strings. The first one is represented 
by the statistics of null detections from pure Gaussian CMB anisotropies, with 
nominal noise and angular resolution of the operating satellites, which, needs 
to be quantified prior to data analysis. Consequently, as the second 
step, one can use the available models and data of Galactic and 
extra-Galactic foreground signal simulations at Planck frequencies, to select 
the sky area where false detections provide a negligible effect with respect to
the main contaminant represented by the background, Gaussian and uniformly 
distributed, CMB anisotropies. In this paper we discuss and specialize the 
technique discussed in \cite{jeongsmoot1} to deal with these issues.
In Section 2 we review in detail the effects of cosmic strings on the CMB
temperature anisotropies and outline the string search algorithm. In Section 3,
we apply the technique to simulated Planck data sets, quantifying the
statistics of null detection based on the nominal performances of the 
satellite, as well as selecting the sky region which is available for searches 
based on the current knowledge of Galactic and extra-Galactic foreground 
emission. Concluding remarks are in Section 4.

\section{Cosmic String effect on CMB and detection algorithm}
\label{effects}
Consider a cosmic string with mass per unit length $\mu$, velocity $\bf{\beta}$
and direction $\hat{\bf{s}}$ both of which are perpendicular to the line of 
sight, back-lighted by a uniform black body radiation background
of temperature $T$. Due to the angular deflection $\delta = 8\pi G\mu$ in the
conic space-time around the string, there is Doppler shift on photons
passing around it on different sides, causing a temperature step across the
string \cite{KaiserStebbins,Gott} with magnitude given by
\begin{equation}\label{eq1}
\label{a}\frac{\delta T}{T} = 8 \pi G \mu \gamma \beta\ ,
\end{equation}
where $\gamma =1/\sqrt{1-\beta^2}$.
In a matter dominated universe the projected angular length of string in the
redshift interval $\fsq{ z_1, z_2}$ scales as $\sqrt{z_1}$ - $\sqrt{z_2}$.
The standard cosmological model ($\Omega_{DE}\sim 0.7,\:\Omega_m\sim 0.3$)
places the decoupling of the CMB, also known as last scattering, at an epoch
which corresponds to a redshift $z_{ls}\sim 1100$. The apparent angular size
of the CMB horizon at last scattering is therefore given by
\begin{equation}\label{eq2}
R = \frac{1}{\sqrt{z_{ls}}}
=1.8^\circ \left( \frac{1000}{z_{ls}} \right)^{1/2}\sim 1.7^{\circ}\ .
\end{equation}
Thus, a convenient choice for the angular size of patches to be examined for
cosmic string signature through (\ref{eq1}) is simply the angle subtended by
the CMB horizon at decoupling, as the maximum extent of the conic space-time
formed around the string would be limited within the horizon at decoupling.
The largest causally connected patch possible would be optimal in this pattern 
search, because a gradient or step-like pattern generated by white noise would 
be suppressed by $\sim 1/\sqrt{N_{\mathrm{pixel}}}$ factor, and our choice the 
size of a test patch as the horizon size at the decoupling can be justified. 
\FIGURE[ht]{
\includegraphics[width=3.5cm,angle=0]{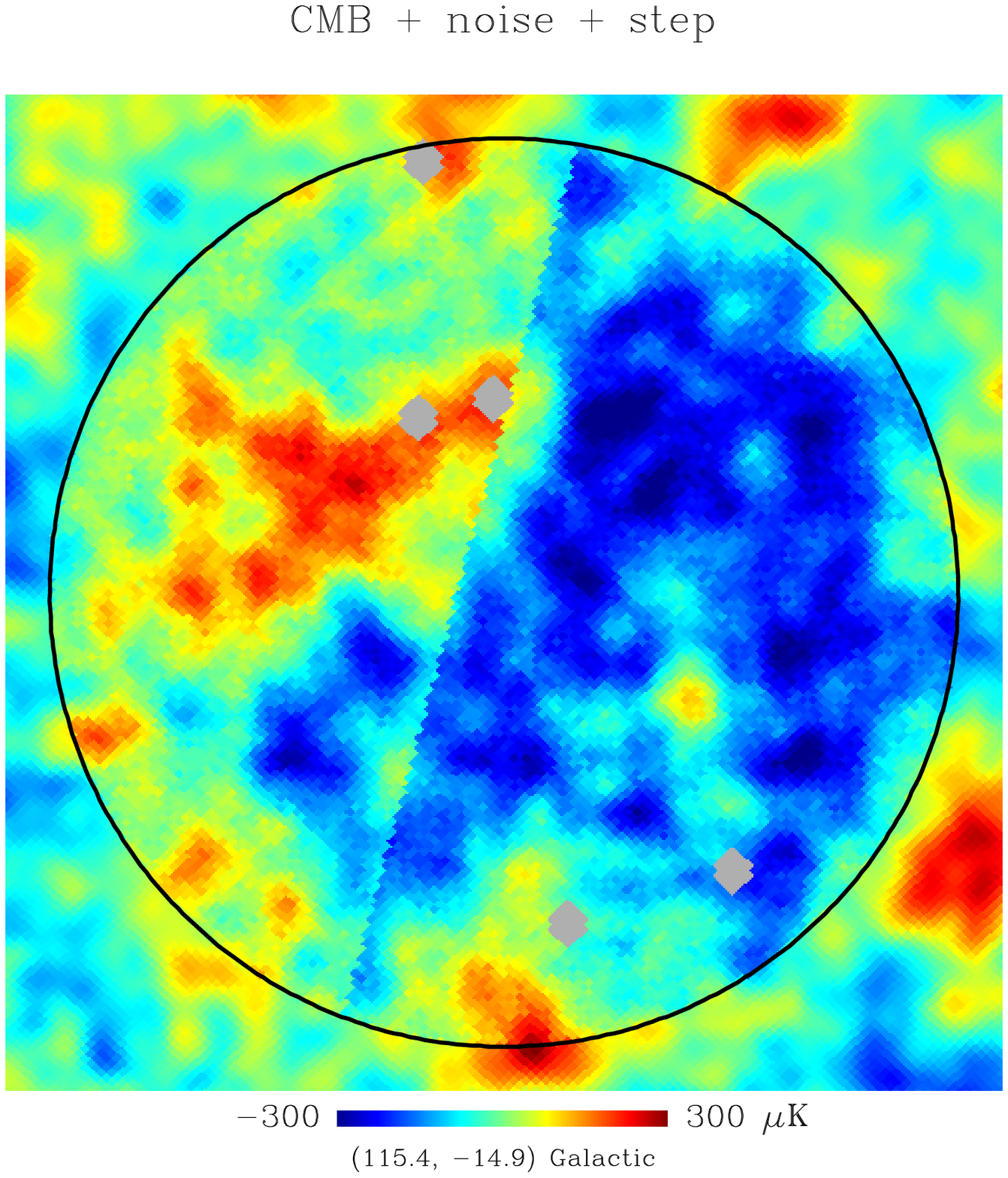}
\includegraphics[width=3.5cm,angle=0]{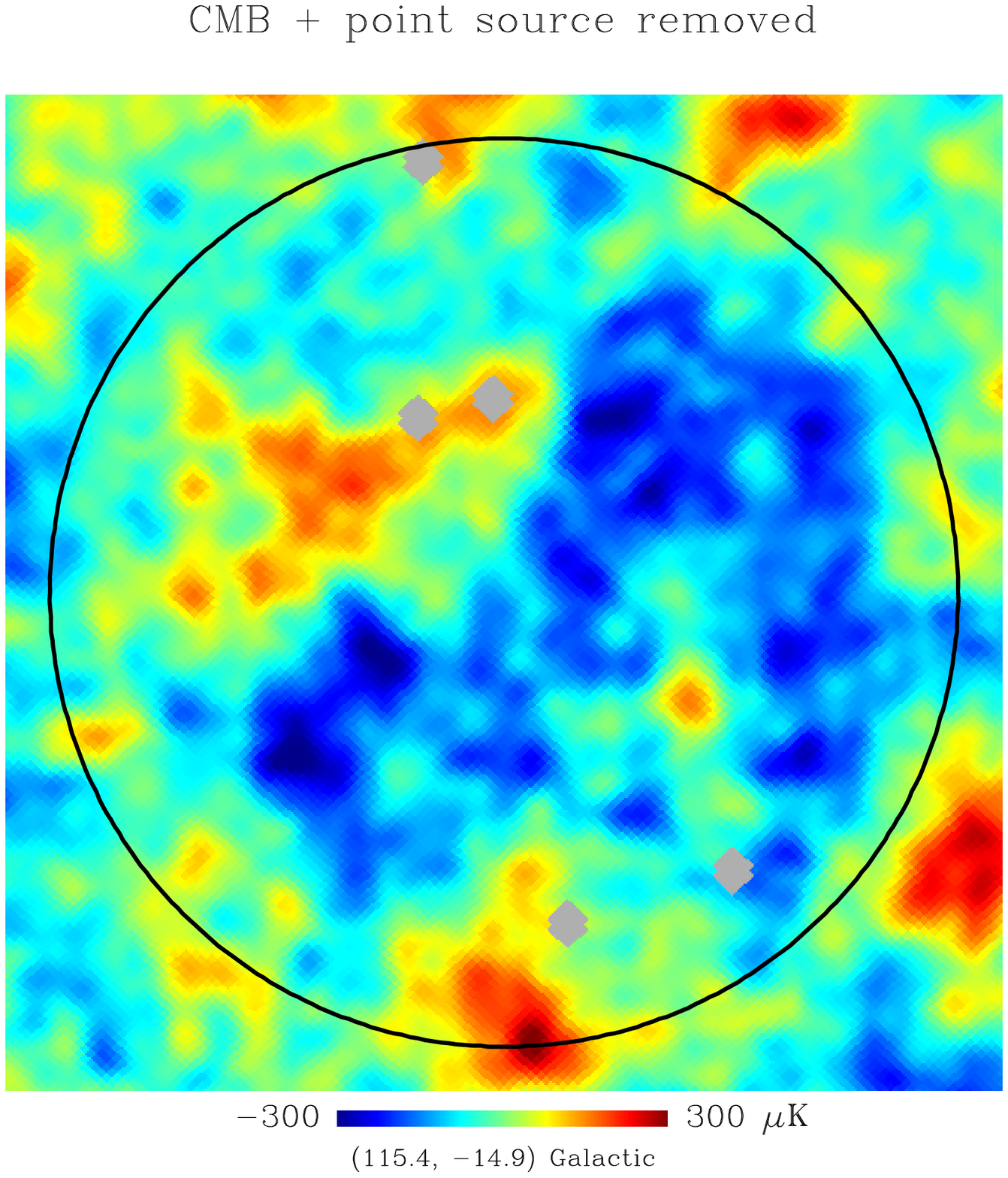}
\includegraphics[width=3.5cm,angle=0]{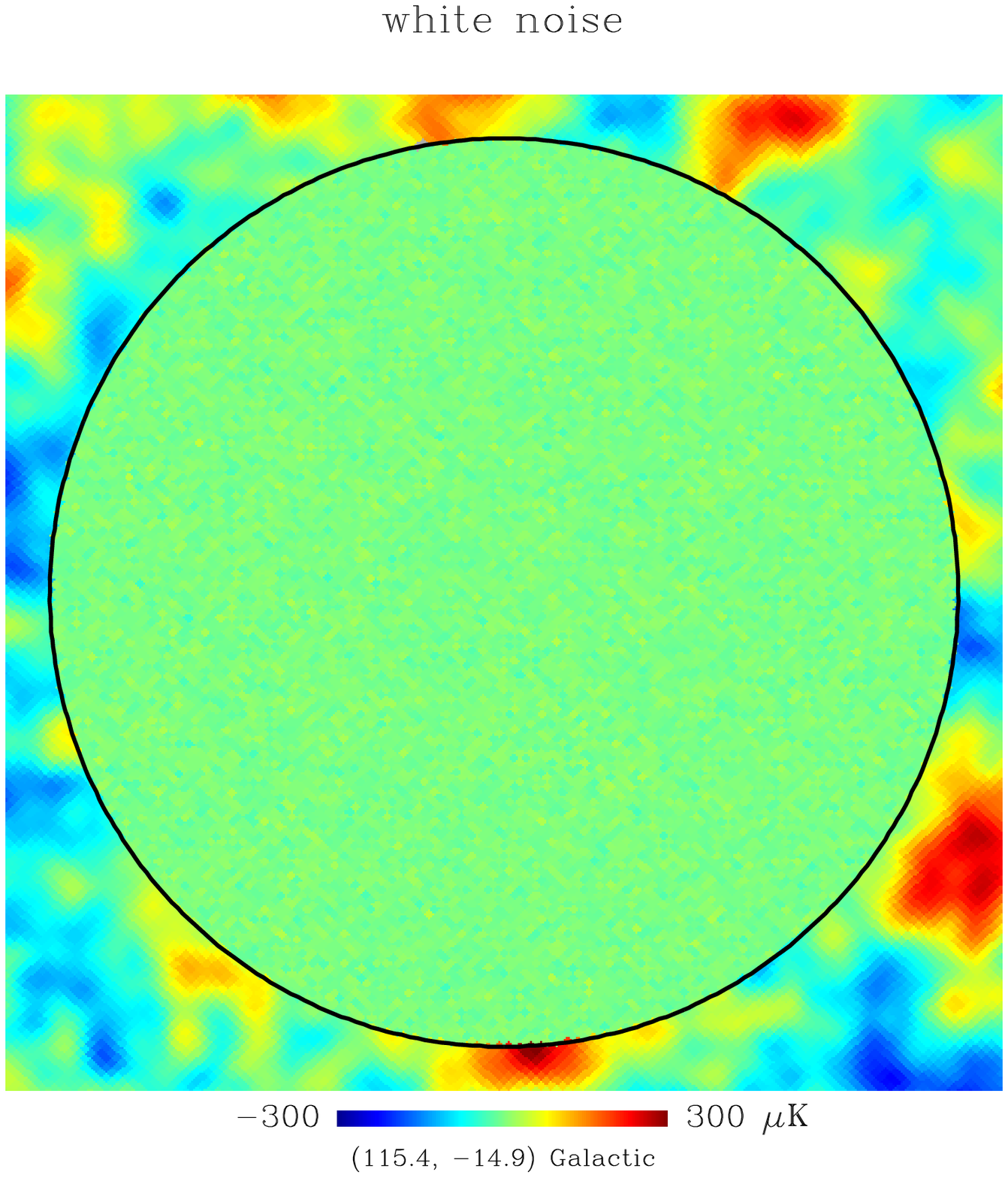}
\includegraphics[width=3.5cm,angle=0]{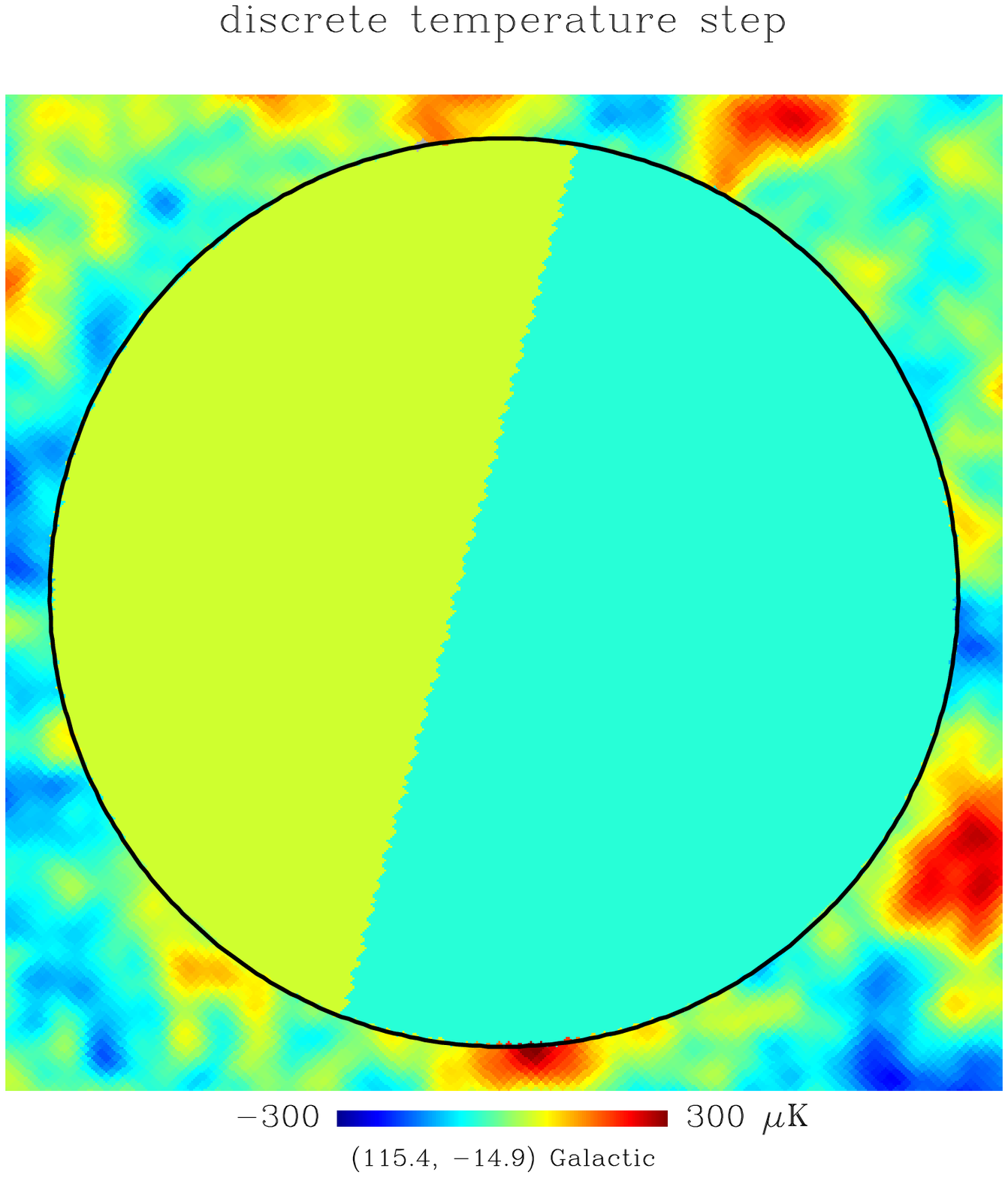}
\caption{\label{fig1}Decomposition of a patch of sky map which contains a 
discrete temperature step. The gray spots represent masked point sources. The 
discrete temperature step is slanted $15^{\circ}$ from the polar axis. The 
instrumental noise was generated with power $\sigma_N=10\muK$ and the height 
of temperature step is $100\muK$.}}
The anisotropy in the CMB temperature sky map is caused by several processes
including intrinsic perturbations in the photon fluid, metric fluctuations, 
possibly topological defects like cosmic strings. CMB photons pick up secondary
anisotropies when traveling across forming structures on the line of sight
toward us. Foreground anisotropies are caused by the emission from
extra-Galactic point sources and our own Galaxy. Finally, the instrument adds
instrumental noise and various systematics. By considering intrinsic CMB
anisotropies, noise and the string contribution, one has
\begin{equation}\label{eq3}
T_{\mathrm{pixel}}=T_{\mathrm{CMB}}+T_{{\mathrm{noise}}}+T_{{\mathrm{step}}}\ .
\end{equation}

We employ five parameters which characterize a circular patch of string
embedded sky map, $T_0$, $\sigma_G$, $\Delta$, $p$ and $\alpha$
\cite{jeongsmoot2}. They represent the background uniform CMB temperature,
the sum of the CMB anisotropy and noise variances, the amplitude of the
discrete temperature step, the ratio of the blue-shifted pixels with respect
to the total, and the orientation of the step, respectively. To recover these
parameters from a given CMB region to be examined, where the individual
components are superimposed, we define five observables which can be expressed
in terms of $\fpr{T_0,\sigma_G,\Delta ,p,\alpha}$:
\begin{eqnarray}
{\mathrm{mean}}\:\mu &\equiv&\frac{1}{\pi R^2}\oint T\id A\label{eq4}\\
{\mathrm{variance}}\:\sigma^2 &\equiv&\frac{1}{\pi R^2}\oint (T-\mu )^2\id A\label{eq5}\\
{\mathrm{dipole\: moment}}\:{\bf{D}} &\equiv&\frac{1}{(\pi R^2)^{3/2}}
\oint T\cdot\bfr\id A\label{eq6}\\
{\mathrm{inertia\: moment}}\: I&\equiv&\frac{1}{(\pi R^2)^2}\oint T\bfr^2\id A
\label{eq7}
\end{eqnarray}
where the integrations are done over the examined sky area, chosen circular
for simplicity, with radius $R$ corresponding to the angular size of the
horizon at decoupling, as we mentioned above.
When a discrete temperature step is present and all the other components
are suppressed, the observables defined in (\ref{eq4}) - (\ref{eq7}) can
be analytically computed and expressed in terms the five step parameters:
\begin{eqnarray}
\mu_{\Delta}&=&\Delta\fpr{p-\frac{1}{2}}\label{eq8}\\
\sigma_{\Delta}^2 &=&\Delta^2p\fpr{1-p}\label{eq9}\\
D_{\Delta} &=&\frac{2\Delta}{3\pi^{3/2}}\fpr{1-\frac{x_c^2}{R^2}}^{3/2}\label{eq10}\\
I_{\Delta} &=&\frac{\Delta}{\pi^2}\fsq{\frac{x_c}{3R}\fpr{1-\frac{x_c^2}{R^2}}^{\frac{3}{2}}+\pi\fpr{p-\frac{1}{2}}}\label{eq11}\
\end{eqnarray}
where the discrete temperature step can be written as
\begin{equation}\label{eq12}
T_{{\mathrm{step}}}=\Delta\fsq{\theta\fpr{x-x_c}-\frac{1}{2}}\ ,
\end{equation}
and $x_c$ ($-R <x_c<R$) is the $x$-coordinate of the step with
$x$-axis running normal to the temperature step, and $\theta\fpr{x-x_c}$
represents a step function. The relation between $p$ (red-shifted area/total
area) and $x_c$(coordinate of discontinuity) is given by
\begin{equation}\label{eq13}
p=\frac{1}{2}-\frac{x_c}{\pi R}\sqrt{1-\frac{x_c^2}{R^2}}-\frac{1}{\pi}\sin^{-1}\fpr{\frac{x_c}{R}}\ .
\end{equation}
The expressions for the observables (\ref{eq8}) - (\ref{eq11})
can be used to find the step parameters by solving for them, recovering
the input ones. Now, taking into account the presence of the non-string
CMB fluctuations and noise, the observables can be calculated to the first
order in the step parameters:
\begin{eqnarray}
\mu &=& T_0+\Delta\fpr{p-\frac{1}{2}}\label{eq14}\\
\sigma^2 &=&\sigma_{\mathrm{CMB}}^2+\sigma_N^2+\Delta^2p\fpr{1-p}\label{eq15}\\
D &=&\fabs{\bf{D}}=\frac{2\Delta}{3\pi^{3/2}}\fpr{1-\frac{x_c^2}{R^2}}^{3/2}\label{eq16}\\
I &=&\frac{T_0}{2\pi}+\frac{\Delta}{\pi^2}\fsq{\frac{x_c}{3R}\fpr{1-\frac{x_c^2}{R^2}}^{\frac{3}{2}}+\frac{\pi p}{2}-\frac{1}{4}}\label{eq17}\ .
\end{eqnarray}
At this point, we solve for the temperature step parameters
($\alpha$ (orientation), $T_0$ (uniform background temperature),
$x_c$ (location of temperature step), $\Delta$ (height of temperature step),
$\sigma_G^2=\sigma_{\mathrm{CMB}}^2+\sigma_N^2$) in terms of observables:
\begin{eqnarray}
\alpha &=& {\mathrm{Im}}\fsq{\ln\fpr{\frac{D_x+iD_y}{D}}}\label{eq18}\\
\frac{x_c}{R} &=& \frac{1}{\sqrt{\pi}}\frac{2\pi I-\mu}{D}\label{eq19}\\
\Delta &=& \frac{2\pi^{3/2}}{3}D\fpr{1-\frac{x_c^2}{R^2}}^{-3/2}\label{eq20}\\
T_0 &=& \mu-\Delta\fpr{p-\frac{1}{2}}\label{eq21}\\
\sigma_G^2&=&\sigma_{\mathrm{CMB}}^2+\sigma_N^2=\sigma^2-\Delta^2p\fpr{1-p}\ .\label{eq22}
\end{eqnarray}
Note that we use $p$ and $x_c$ interchangeably since they are related by
(\ref{eq13}). Expressions (\ref{eq18}) - (\ref{eq22}) are the key formulas
in the algorithm used to compute the step parameters in any give horizon sized
circular patch of the sky. Due to the presence of CMB and noise fluctuations
(and foreground emissions, to be considered next), the results will have
statistical fluctuations. The dispersions of the outputs are the estimate for
the error associated to the quantities measured by the algorithm discussed
here.
\FIGURE[ht]{
\includegraphics[width=7.0cm,angle=0]{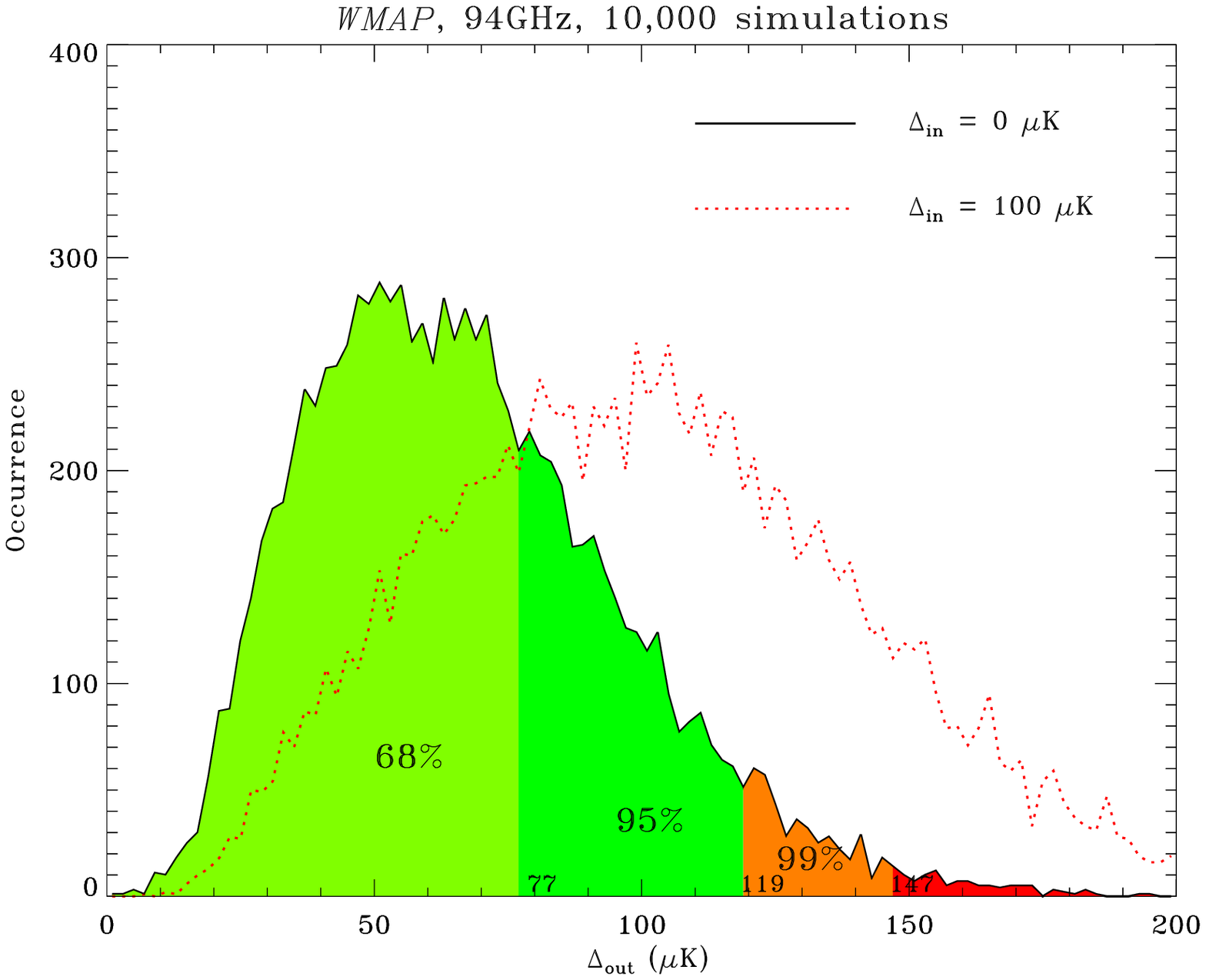}
\includegraphics[width=7.0cm,angle=0]{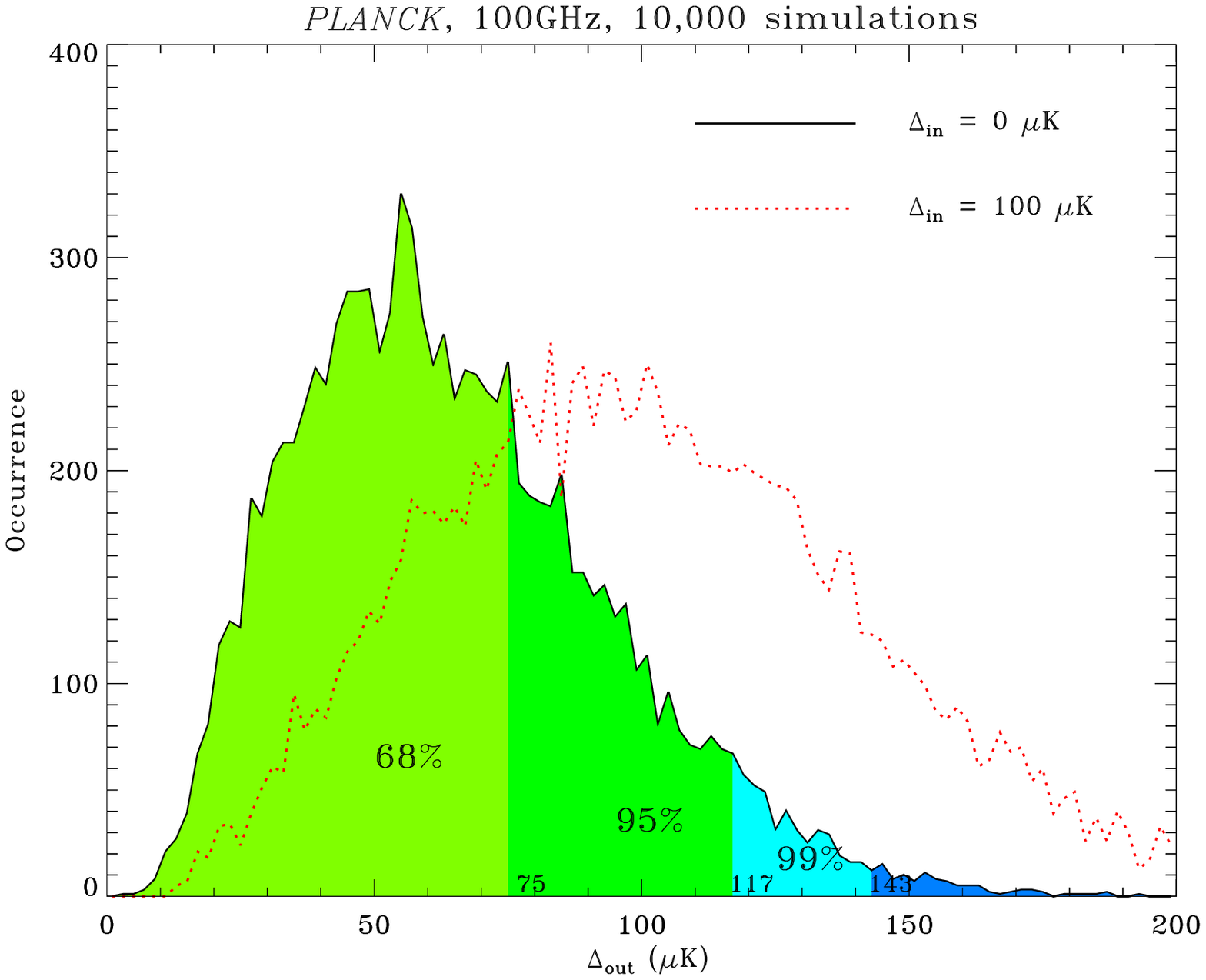}
\caption{\label{f2}Simulation results with \WMAP (left) and \PLANCK (right)
level simulated CMB temperature maps. Red-dotted curves are results with
$\Delta_{\mathrm{in}}=100\muK$.}}

\TABLE[ht]{
\begin{tabular}{lccc||ccc}
\hline\hline
{$\: $}&\multicolumn{3}{c||}{$C_{l\le 140}$, $\sigma_N=10\muK$}
&\multicolumn{3}{c}{$C_{l>140}$, $\sigma_N=10\muK$}\\
\hline
Cumulative occurrence&68\%&95\%&99\%&68\%&95\%&99\%\\
\hline
$N_{\mathrm{side}}=2^9$&67&109&139&29&41&51\\
$N_{\mathrm{side}}=2^{10}$&67&109&141&29&41&49\\
$N_{\mathrm{side}}=2^{11}$&69&113&141&27&41&49\\
\hline\hline
{$\: $}&\multicolumn{3}{c||}{$C_{l\le 140}$, $\sigma_N=100\muK$}
&\multicolumn{3}{c}{$C_{>140}$, $\sigma_N=100\muK$}\\
\hline
Cumulative occurrence&68\%&95\%&99\%&68\%&95\%&99\%\\
\hline
$N_{\mathrm{side}}=2^9$&71&109&135&35&49&59\\
$N_{\mathrm{side}}=2^{10}$&67&109&135&29&43&51\\
$N_{\mathrm{side}}=2^{11}$&67&109&131&29&41&49\\
\hline\hline
\end{tabular}
\caption{\label{table1}Comparison of cumulative occurrence levels of confusion
outcomes for large and small scale CMB anisotropy power, with varying noise
amplitude, map resolution.}}
\section{String detection statistics and foreground effects for satellite measurements}
In this section, we apply the search algorithm outlined in the previous
section in order to examine two important aspects of string search for
satellite CMB measurements. The latter are characterized by a wide sky area
and frequency coverage, mapping in particular the Galactic foreground
contamination to high accuracy in bands where it is dominant, as well as
high angular resolution and sensitivity. When looking for string signatures,
one has to characterize the statistics of null detections caused by a pure
Gaussian spectrum of CMB anisotropies, in order to assess the probability that
a given event is actually caused by a string and not by an ordinary CMB
fluctuations; moreover, the spurious detections caused by foreground emissions
have to be quantified and controlled. In this Section we address both aspects.
\subsection{Null detections from Gaussian CMB anisotropies}
We use the publicly available simulation codes CAMB\footnote{http://camb.info/}
and HEALPix\footnote{http://healpix.jpl.nasa.gov} \cite{gorski.et.al}
to generate simulation maps with a consensus $\Lambda$ Cold Dark Matter 
($\Lambda$CDM) cosmology input parameters ($\Omega_b=0.046$, $\Omega_{\mathrm{CDM}}=0.233$, $\Omega_{\Lambda}=0.721$, $h=0.701$) (see \cite{larson.et.al} for 
updated results on cosmological parameters). We run the detection algorithm 
described in the previous Section on these maps, and evaluate the statistics 
of null detections, adopting the nominal performances of the currently 
operating \WMAP and \PLANCK satellites. 
Indeed, we do not expect significant differences
in the two cases, as the signal we are looking for is mainly concentrated
on the degree scale, where both instruments are substantially limited by
cosmic variance. The results are shown in Figure \ref{f2}, solid line with
colored integral. On the left, the angular resolution and nominal sensitivity
were set corresponding to the \WMAP W band with central frequency at 94 GHz,
FWHM=13.2 arc minutes, $N_{\mathrm{side}}=2^9$, white noise noise
rms $=\sigma_N=35\muK$ on $0.3^{\circ}\times 0.3^{\circ}$. On the right, the
angular resolution and sensitivities were chosen correspondingly to the
\PLANCK 100 GHz channel on-board the High Frequency Instrument, with FWHM=10
arcmin, $N_{\mathrm{side}}=2^{11}$, noise rms $=\sigma_N=2.5\muK$ on squared
pixels with side equal to the FWHM. The first thing to note is that
the distributions of the null detected $\Delta^2_{\mathrm{out}}$ are
similar as we anticipated, and approximately follow a $\chi^2$-distribution,
due to the randomness of Gaussian peaks which determine the spurious
signal. The color filled curves in Figure \ref{f2} show the different
confidence level that a given event in a real measurement is not due
to spurious detection from Gaussian anisotropies. For \WMAP,
68\% of occurrences falls in $0<\Delta <77\muK$, 95\% in
$\Delta <119\muK$ and 99\% in $\Delta <147\muK$, respectively.
For \PLANCK, we have 68\% in $0<\Delta <75\muK$, 95\% in $\Delta <117\muK$ and
99\% at $\Delta <143\muK$. The dotted lines also show the cases in which the
sky signal was added with a step with $\Delta_{\mathrm{in}}=100\muK$ in
each area examined, clearly indicating that the search algorithm is working,
as the distribution peaks at the chosen value of $\Delta_{\mathrm{in}}$.
The fact that \WMAP and \PLANCK show nearly same results indicates that the
effects of map resolution or instrumental noise level applied here on the
result are nearly negligible compared to other factors, which we will discuss
further in the following. \\
In conclusion of this Section, we show that the main contaminant for cosmic
string search with specifications corresponding to the operating satellites
are not given by the noise but by the underlying Gaussian fluctuations, which
are dominated by acoustic oscillations on the degree scale which we are
focusing on. In order to contrast the effects of super-horizon scale
fluctuations and sub-horizon ones, we compute the statistics of null
detections by considering simulated maps with anisotropy power on the
super-degree angular scales ($l\le 140$) and sub-degree ones ($l\ge 140$),
separately. Notice that the rms fluctuation power is about the same for the
two sets:
\begin{equation}\label{eq23}
\sum_{l=2}^{140}\frac{2l+1}{4\pi}C_l=\sum_{l>140}\frac{2l+1}{4\pi}C_l\ .
\end{equation}
In Table \ref{table1}, one sees the variation of the confidence levels
for the large and small scale power as a function of map resolution,
noise level. At the first glance, it is clear that, while noise
and map resolution do not induce significant differences, the most
significant improvement is obtained when super-degree and degree
scale power is taken out.

%%%%\TABLE[th]{
\begin{table*}
\begin{center}
\begin{tabular}{cccc}
\hline\hline
 &\multicolumn{3}{c}{Cumulative occurrence ($\muK$)}\\
\hline
Frequency&68\%&95\%&99\%\\
\hline
30 GHz&69&109&133\\
44 GHz&73&115&141\\
70 GHz&75&117&143\\
100 GHz&77&119&147\\
143 GHz&77&119&145\\
217 GHz&77&119&147\\
353 GHz&77&119&147\\
\hline\hline
\end{tabular}
\caption{\label{table2}Cumulative occurrences of string signal detection from 
the simulated full sky CMB maps of 7 \PLANCK frequencies. For each frequency 
band, smoothing and instrumental noises are applied accordingly as specified 
in \cite{planckbluebook}.}  %%}
\end{center}
\end{table*}

\subsection{Spurious string signals from foregrounds}
\FIGURE[ht]{
\includegraphics[width=4.5cm,angle=90]{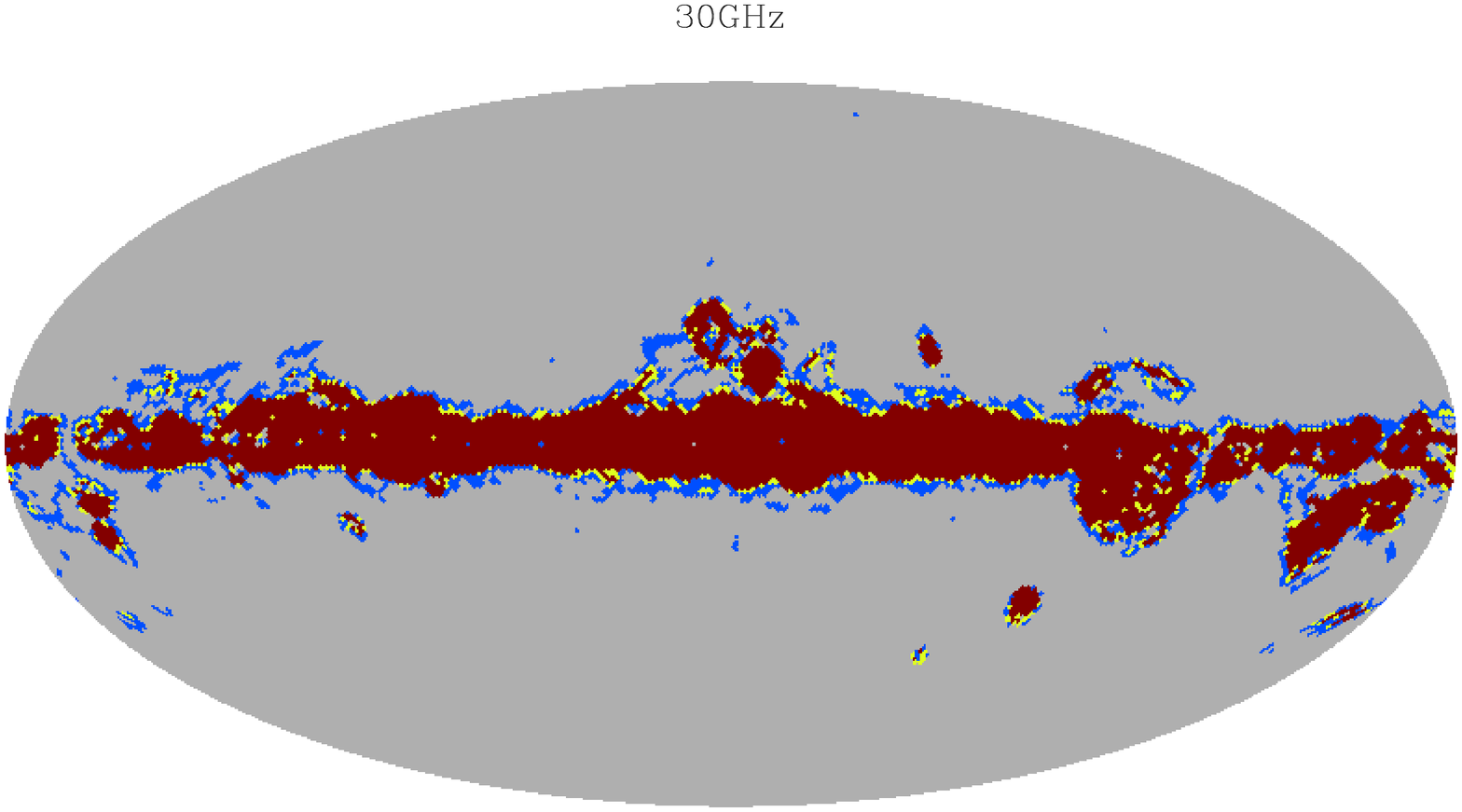}
\includegraphics[width=4.5cm,angle=90]{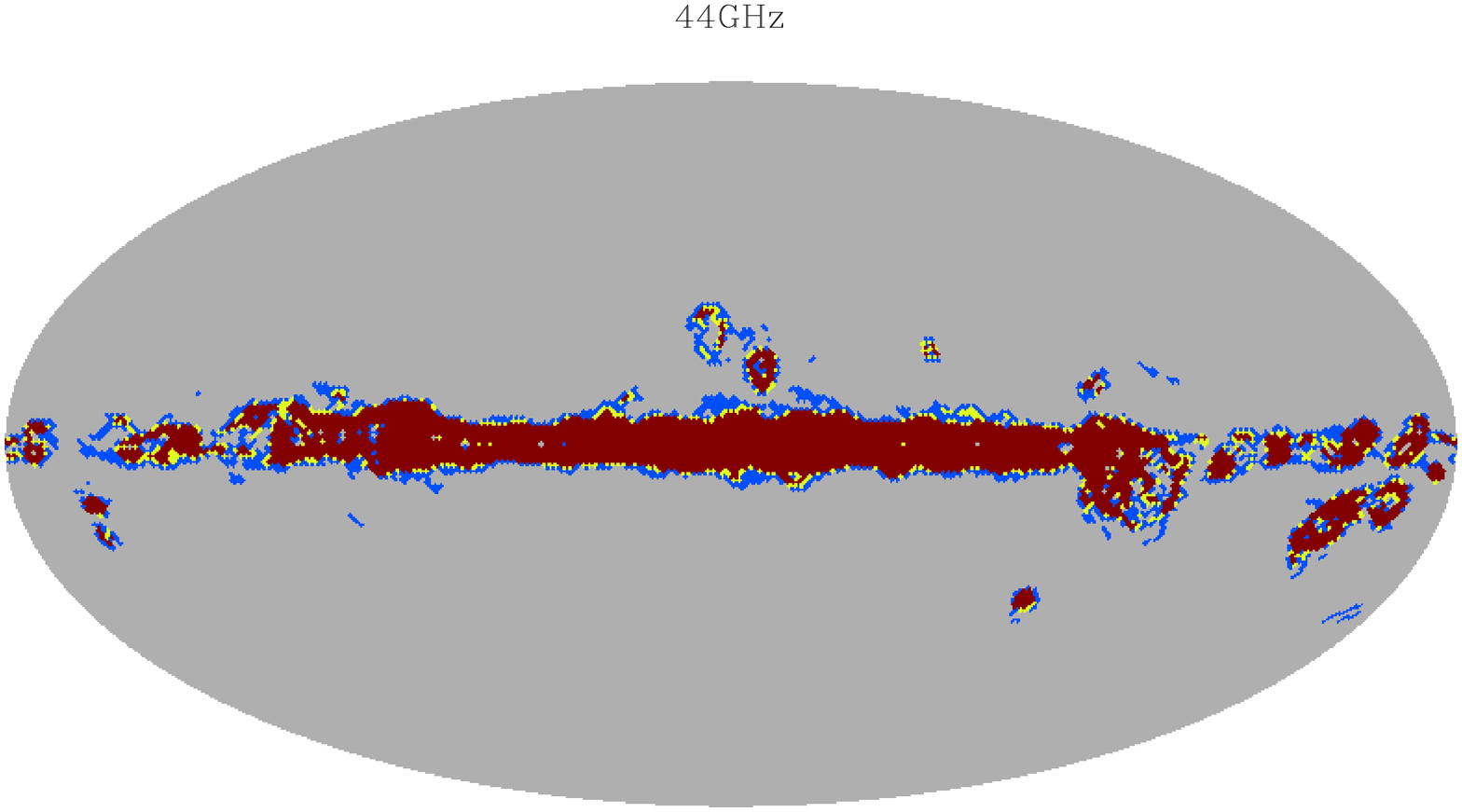}
\includegraphics[width=4.5cm,angle=90]{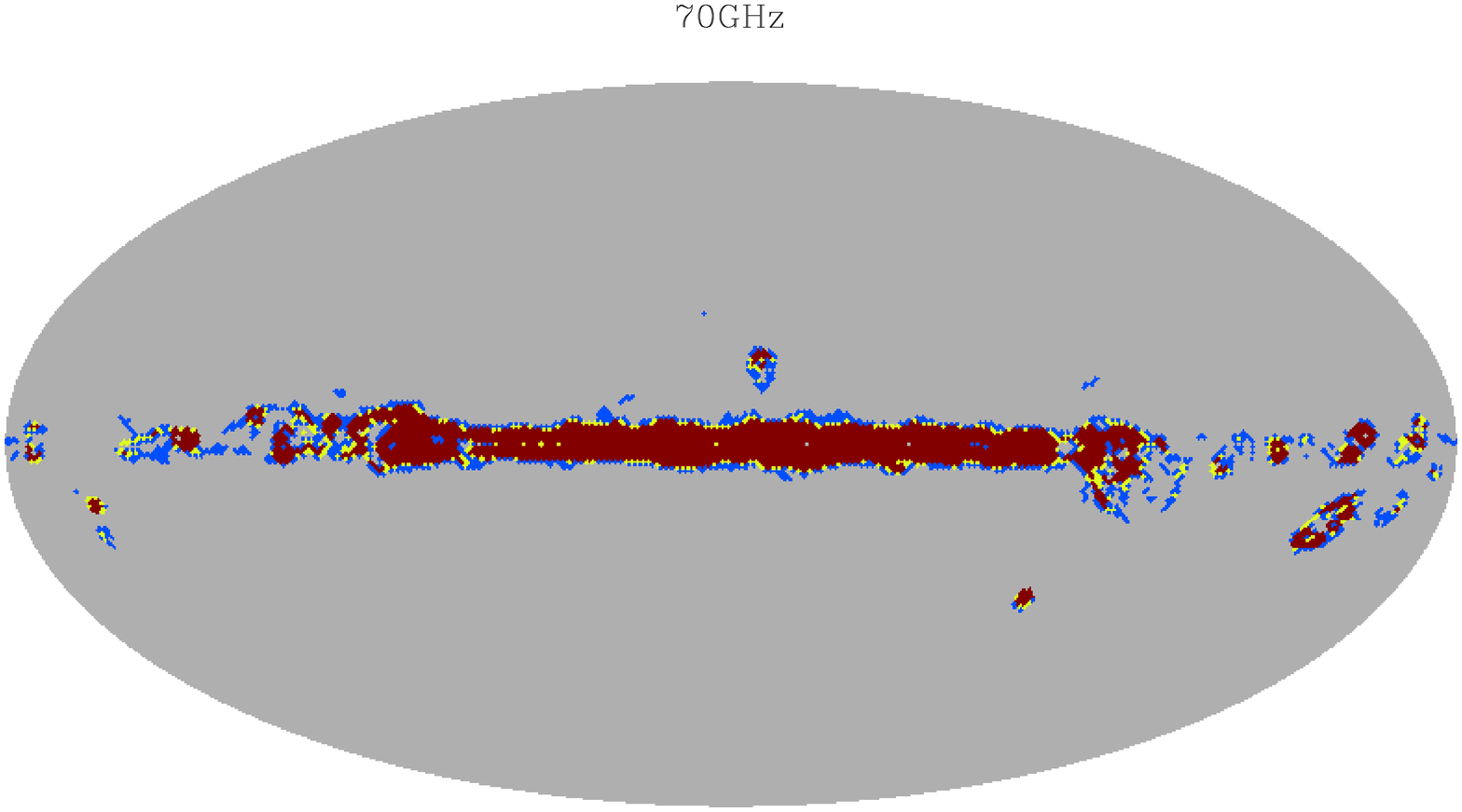}
\includegraphics[width=4.5cm,angle=90]{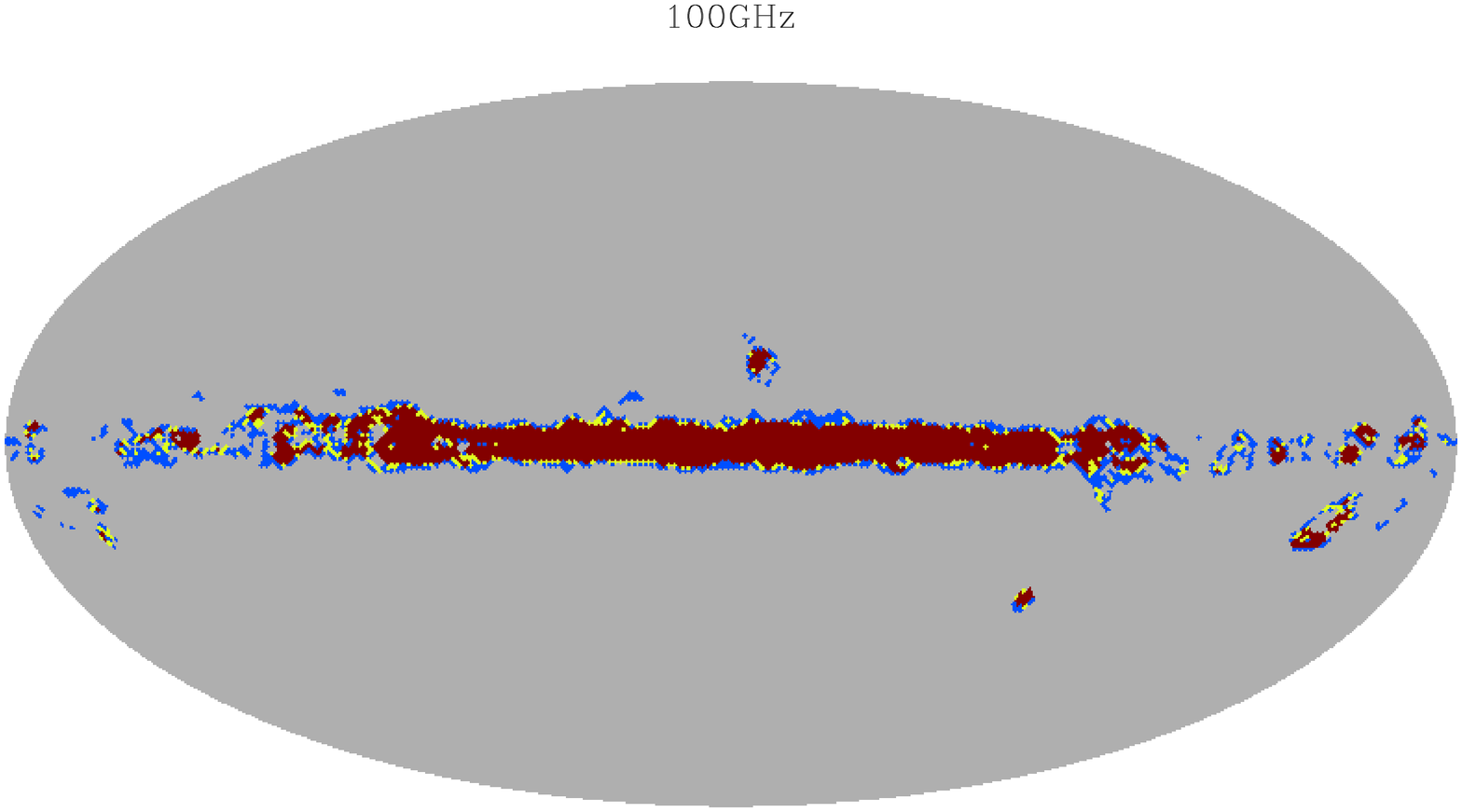}
\includegraphics[width=4.5cm,angle=90]{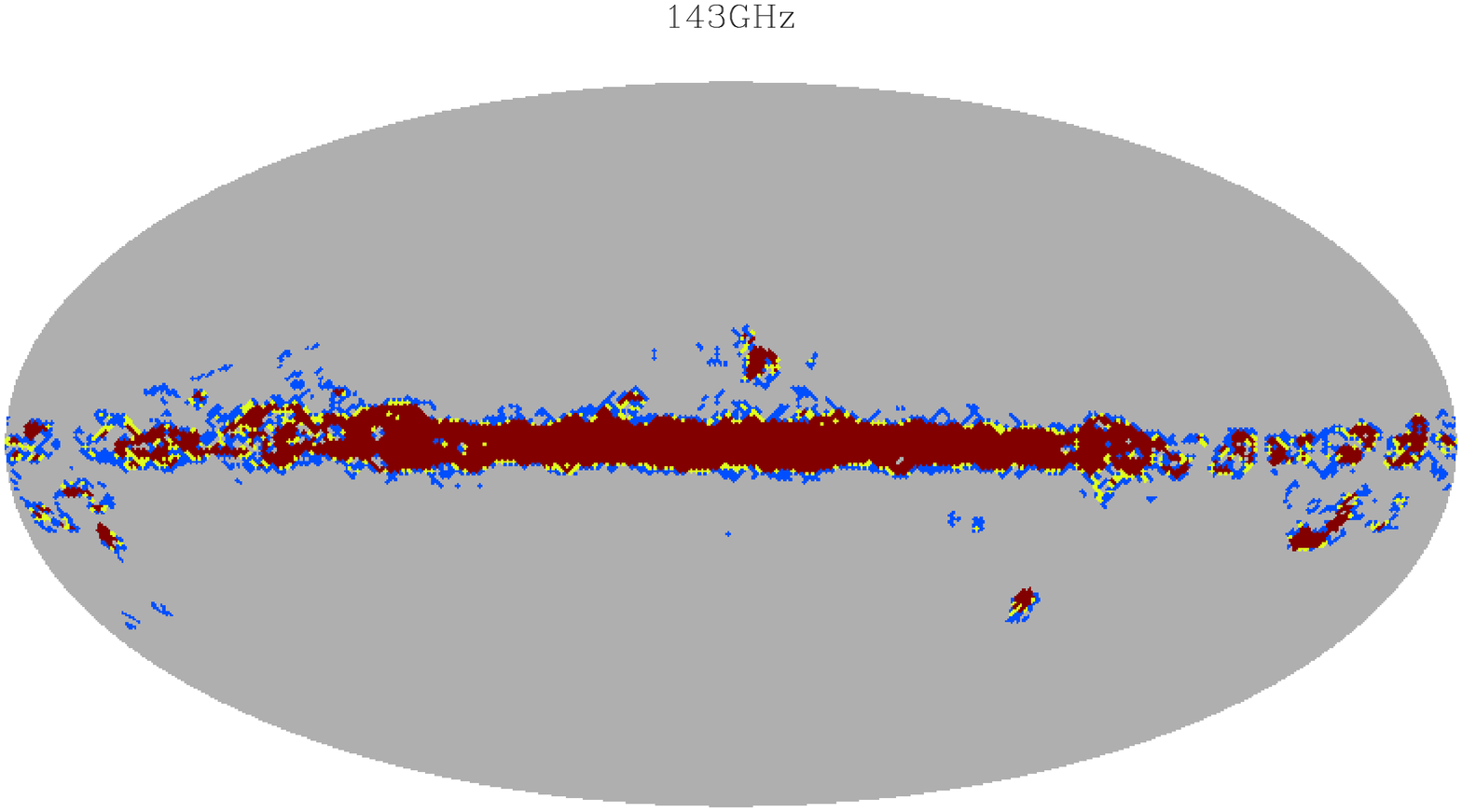}
\includegraphics[width=4.5cm,angle=90]{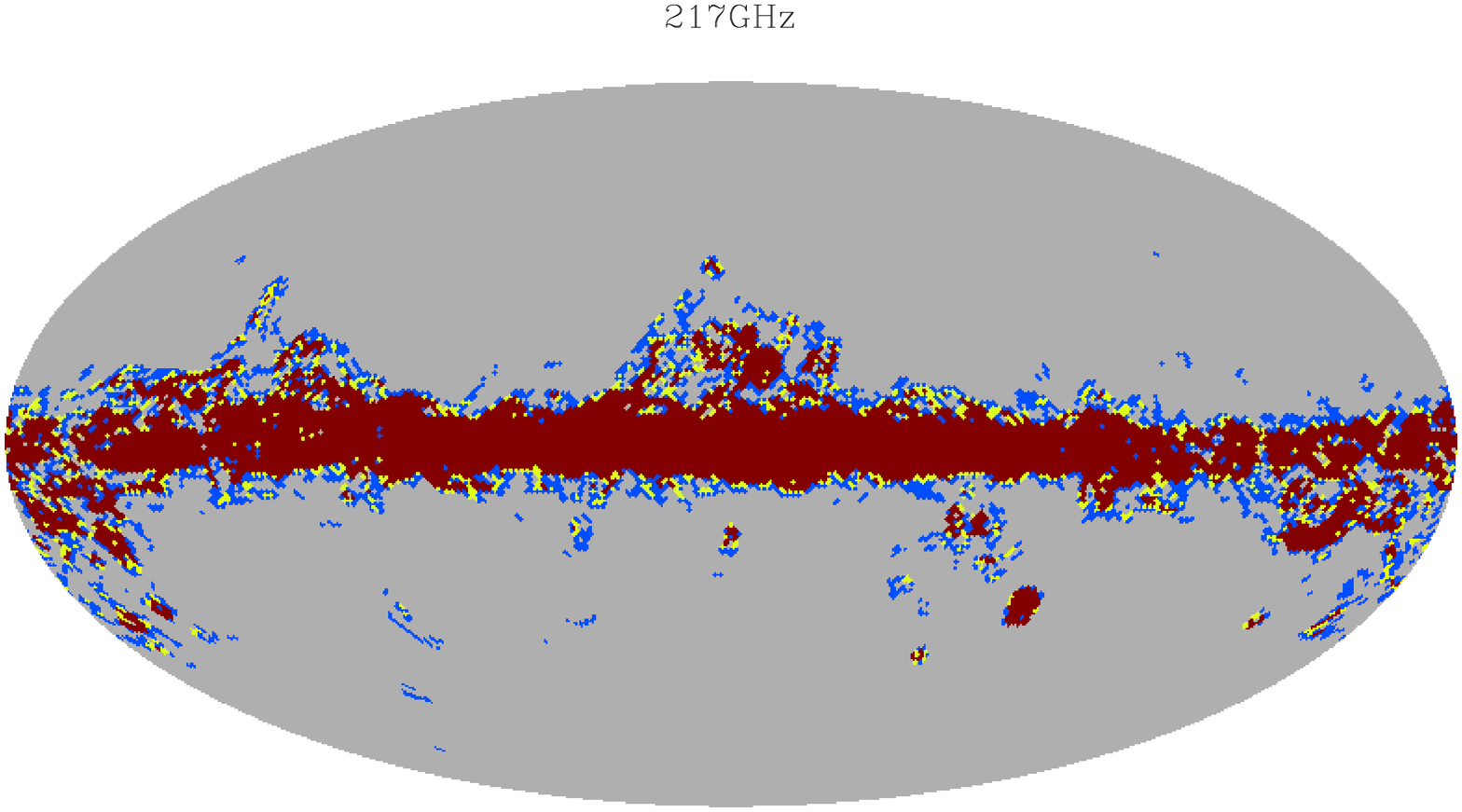}
\includegraphics[width=4.5cm,angle=90]{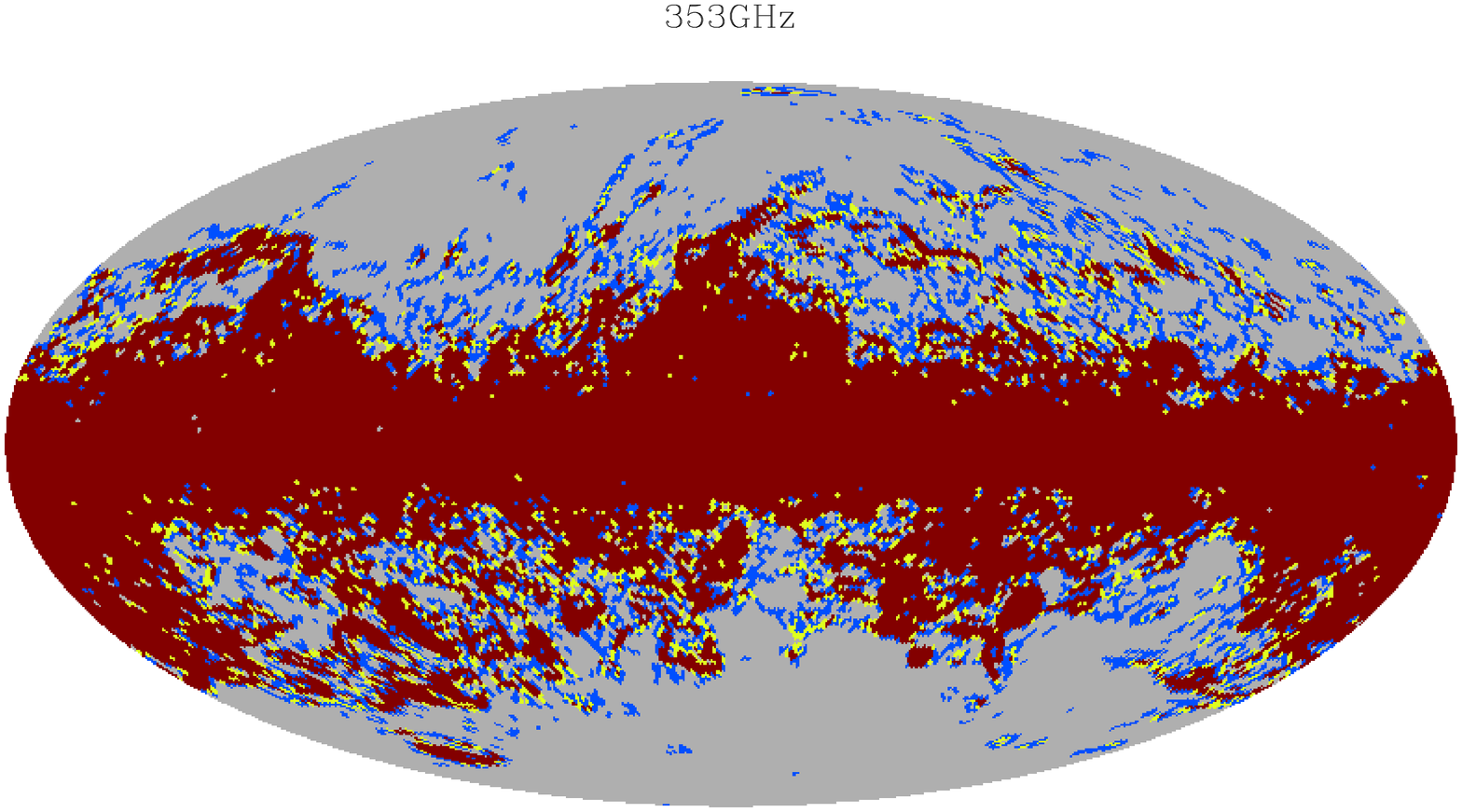}
\caption{\label{f3}Area overshadowed by strong signals from diffuse
foreground emissions for each of the 7 \PLANCK frequencies considered,
30, 44, 70, 100, 143, 217 and 353 GHz. Blue, yellow and red colored areas
denote the regions where the confusion outcomes by foreground emissions exceed
68\%, 95\% and 99\% of cumulative occurrences from simulated CMB maps
for each corresponding frequencies, which are shown in Table \ref{table2}.}}
Foregrounds are considered as one of the ultimate limitation of CMB
measurements, and represent a major source of contamination, in particular
for cosmic string searches. Multi-frequency data analysis techniques are being
studied for \PLANCK \cite{leach.et.al}, and applied successfully
to the \WMAP data \cite{dunkley.et.al} in order to separate them from the
main CMB emission. Here we take a conservative
approach, by exploiting the current foreground observations outside
and inside the microwave band, performing an all sky analysis looking
for regions where they cause spurious string detections, focusing on
the \PLANCK frequency channels going from 30 to 353 GHz. \\
Let us consider extra-Galactic point sources first. The usual
technique to identify and remove them is to isolate $~5\sigma$ events
above the CMB and noise standard deviations, and removing them with
optimal filters \cite{herranz.et.al}.
Residual sources behave as a correlated noise due to beam smoothing
which remains in the maps. We simulated randomly scattered point sources
according to the existing modeling of the radio and infrared populations
\cite{dezotti.et.al, gonzaleznuevo.et.al}. After removing the $~5\sigma$
signals, causing typically a few holes of beam size in the patches we 
consider, we run our cosmic string search algorithm to look for spurious
detections. By comparing the recovered spurious $\Delta$, we find that no 
significant disturbance from unresolved point sources is observed, due to 
their sub-dominance with respect to the CMB acoustic oscillations, which 
largely prevails as a contaminant.\\
The situation is markedly different for the diffuse foregrounds coming
from our own Galaxy. We construct all sky maps including the effect of
three of the main diffuse Galactic emission mechanisms:
the synchrotron emission is caused by electrons spiraling around the lines
of the Galactic magnetic field, using the data by \cite{haslam.et.al},
treated by \cite{giardino.et.al} with spectral index obtained
from WMAP analysis \cite{bennett.et.al}, see for more
detail{\footnote{ftp://ftp.rssd.esa.int/pub/synchrotron/README.html}};
the dust total intensity is based on the analysis of IRAS and DIRBE data
by \cite{finkbeiner.et.al}, implementing model 8 of frequency scaling
including spatial variations of dust frequency scaling; the free-free emission 
from Brehmstraahlung of electrons hitting ions in the Galaxy, traced by 
H$\alpha$ emission, has also been
included{\footnote{http://astrometry.fas.harvard.edu/skymaps}}.\\
We simulate sky maps with these emissions included at 30, 44, 70, 100, 143, 217
and 353 GHz, and apply the cosmic string detection algorithm. As expected,
we do find regions with high rate of contamination, concentrated close to the 
Galactic plane, at all frequencies. We show the results in Figure \ref{f3}, 
where we show all together the regions yielding spurious cosmic string signals
with various confidence levels based on the distribution of null detection
from CMB emission only, outlined in Table \ref{table2}.
Notice that, in addition to the Galactic plane signals, several emissions
are noticeable from intensely emitting Galactic clouds at intermediate
latitudes. Cosmic string searches based on the algorithm presented here should
be avoided in the highlighted regions.

\section{Conclusions}
We studied a direct cosmic string search based on the Cosmic Microwave 
Background (CMB) anisotropies in total intensity, in the context of the 
existing satellite observations, focusing on the case of the ongoing \PLANCK 
survey. The search algorithm uses an algebraic method to detect step like 
structures in CMB maps, caused by the Kaiser-Stebbins effect induced by strings 
moving on the orthogonal plane with respect to the line of sight. We quantified 
two important aspects towards the application of this algorithm to the actual 
data.\\
First, we derived the statistics of null detections due to instrumental
noise and purely Gaussian CMB anisotropies as predicted within the 
$\Lambda$ Cold Dark Matter ($\Lambda$CDM) cosmology. We have shown that the 
biggest cause of confusion for the detection algorithm is caused by the CMB 
acoustic oscillations at wavelengths comparable to the size of the horizon at
decoupling, while noise and angular resolution which are typical of the
operating satellites, are less important. The distribution of null detection
converges to a $\chi^{2}$-distribution. We derived the confidence levels of 
detection of cosmic strings in \PLANCK maps by estimating the 68\%, 95\% and 
99\% of cumulative occurrence of null detection, thresholds give in 
Table \ref{table2}, which corresponds to $G\mu\sim 1.5\times 10^{-6}$ for 
95\% of occurrences in terms of cosmic string tension, according to the 
relation between the string tension and temperature shift of CMB photons in 
\ref{eq1}.\\
On the basis of these results, we were able to evaluate the effect of 
extra-Galactic and Galactic foreground emissions on cosmic string searches, 
by comparison the foreground induced string signals with the detectability 
threshold found in the first part of this work. As expected, we find that after 
removal of the brightest sources exceeding $5\sigma$ times the CMB rms, 
residual confusion due to unresolved point sources does not have a significant 
impact on the string detection algorithm. On the other hand, the diffuse 
Galactic emission from our own Galaxy do cause false detection in a significant
fraction of the sky, not because of their overall intensity, but rather due to 
their gradients, reaching intermediate Galactic latitudes. By using the current
data and models of the diffuse Galactic emissions based on off microwave band 
data as well as the measurements of the \WMAP satellite, we determine the area 
of the sky where false detection are expected due to the Galactic emission, at 
the level of 68\%, 95\% and 99\% of cumulative occurrences for each of seven 
\PLANCK frequencies, where the thresholds are given in Table \ref{table2}. We 
specialize the discussion to seven frequency channels on-board \PLANCK, 
constructing for each one the sky mask to be used for string searches. 
The maximum available area in which the Galactic contamination is sub-dominant 
with respect to the contribution from Gaussian CMB fluctuations is found in the
frequency interval 70-100 GHz, and represents about 7\% of the entire sky. 
These criteria and findings can be used in the actual application of string 
searches in CMB data.\\
For the future work on the pattern search of cosmic strings in CMB temperature 
data, it is desirable to exploit the correlated signals of discrete steps 
lined up close to each other, since the pattern search is mainly pointed at 
the long strings of cosmological scale rather than isolated segments or loops 
of cosmic strings. The methology provided in this paper on finding step signal 
by a cosmic string segment will be the basis for the long string search. 

\subsection{Acknowledgments}
We thank Reno Mandolesi for reading the manuscript and providing useful 
feedback, and Dr. Joaquin Gonzalez Nuevo-Gonzalez for valuable advices and 
discussion on point sources. We acknowledge the WMAP 7-year data which were 
processed and released by LAMBDA\footnote{http://lambda.gsfc.nasa.gov/}.
Some of the results in this paper have been derived using the
HEALPix\footnote{http://healpix.jpl.nasa.gov/}.

\end{document}